\documentclass{article} 
\usepackage{spconf}
\usepackage[utf8]{inputenc} 
\usepackage[T1]{fontenc}    
\usepackage{hyperref}       
\usepackage{url}            
\usepackage{booktabs}       
\usepackage{amsfonts}       
\usepackage{nicefrac}       
\usepackage{microtype}      
\usepackage{xcolor}         

\usepackage{graphicx}
\usepackage{makecell}
\usepackage{multirow}
\usepackage{multicol}
\usepackage[frozencache=true,cachedir=minted-cache]{minted} 
\usepackage{tablefootnote}

\title{TorchAudio: building blocks for audio and speech processing }

\name{
  \begin{tabular}{c}
  \it Yao-Yuan Yang$^{*\dagger 2}$, Moto Hira$^{*1}$, Zhaoheng Ni$^{*1}$, Anjali Chourdia$^1$, Artyom Astafurov$^1$,\\
  Caroline Chen$^1$,
  Ching-Feng Yeh$^1$,
  Christian Puhrsch$^{1}$,
  David Pollack$^3$,
  Dmitriy Genzel$^1$,\\
  Donny Greenberg$^{1}$,
  Edward Z. Yang$^1$,
  Jason Lian$^{\dagger 1}$,
  Jay Mahadeokar$^1$,
  Jeff Hwang$^1$,\\
  Ji Chen$^1$,
  Peter Goldsborough$^4$,
  Prabhat Roy$^1$,
  Sean Narenthiran$^{5}$,
  Shinji Watanabe$^6$,\\
  Soumith Chintala$^{1}$,
  Vincent Quenneville-Bélair$^{\dagger 1}$,
  Yangyang Shi$^{1}$
  \end{tabular}
  \thanks{$^*$ Equal contribution on the writing of this paper. The rest of authors are in alphabetical order.}
  \thanks{$^\dagger$ Work done while at Meta AI.}
}

\address{
  $^1$Meta AI,
  $^2$University of California, San Diego,
  $^3$Solvemate Gmbh,\\
  $^4$Anduril Industries,
  $^5$Grid AI Labs,
  $^6$Carnegie Mellon University
}

\newcommand{\code}[1]{\texttt{#1}}
\newcommand{\package}[1]{\texttt{#1}}

\begin{document}
\ninept   
\maketitle

\begin{abstract}
This document describes version 0.10 of \package{TorchAudio}:
building blocks for machine learning applications in the audio and speech processing domain.
The objective of \package{TorchAudio} is to accelerate the development
and deployment of machine learning applications for researchers and engineers
by providing off-the-shelf building blocks.
The building blocks are designed to be GPU-compatible, automatically differentiable, and production-ready. 
\package{TorchAudio} can be easily installed from Python Package Index repository and
the source code is publicly available under a BSD-2-Clause License
(as of September 2021) at \url{https://github.com/pytorch/audio}.
In this document, we provide an overview of the design principles, functionalities, and 
benchmarks of \package{TorchAudio}.
We also benchmark our implementation of several audio and speech operations and models. 
We verify through the benchmarks that our implementations of various operations and models are valid and perform similarly to other publicly available implementations.
\end{abstract}


\section{Introduction}

In recent years, the usage of open-source toolkits for the development and deployment
of state-of-the-art machine learning applications has grown rapidly.
General-purpose open-source toolkits such as
\package{TensorFlow}~\cite{tensorflow2015-whitepaper}
and \package{PyTorch}~\cite{NEURIPS2019_9015}
are used extensively.
However, building applications for different domains requires additional domain-specific functionality.
To accelerate development,
we established the \package{TorchAudio} toolkit, which provides building
blocks for machine learning applications in the audio and speech domain.

To provide building blocks for modern machine learning applications in the
audio/speech domain, we aim to enable the following three important properties
for each building block:
(1) GPU compute capability,
(2) automatic differentiablility,
(3) production readiness.
GPU compute capability enables the acceleration of the training
and inference process.
Automatic differentiation allows to directly incorporate these functionalities into neural networks enabling full end-to-end learning.
Production readiness means that the trained models can be easily
ported to various environments including mobile devices running
Android and iOS platforms. 

The stability of the toolkit is our top priority.
Our goal is not to include all state-of-the-art technologies, but rather to provide high-quality canonical implementations that users can build upon.
The functionalities provided in \package{TorchAudio} include
basic input/output of audio files,
audio/speech datasets,
audio/speech operations, and
implementations of canonical machine learning models.
To the best of our knowledge,
\package{TorchAudio} is the only toolkit that
has building blocks that span these four functionalities, with the majority satisfying the three aforementioned properties.

To understand how our implementation
compares with others, we conduct an empirical study.
We benchmark five audio/speech operations and three machine learning models
and compare them with publicly available reference implementations.
For audio/speech operations, we find that our implementations achieves or exceed parity in run time performance.
For the machine learning models, we show that our implementations achieve or exceed parity in output quality.

\package{TorchAudio} is designed to facilitate new projects in the audio/speech domain.
We have built up a thriving community on Github with many users and developers.
As of September 2021, we have addressed more than $400$ issues and
merged more than $1250$ pull requests.
We have more than $40$ users that contributed more than $100$ lines of code,
with many of these users being external contributors.
We also observe extensive usage of \package{TorchAudio} in the
open source community.
Currently, there are more than $350$ projects that are forked from \package{TorchAudio} and
more than $5,420$ public repositories on Github that depend on \package{TorchAudio}.

This paper is organized as follows.
We start with reviewing existing open source audio/speech toolkits in Section~\ref{sec:rel}.
Next, we talk about \package{TorchAudio}'s design principles in Section~\ref{sec:design}, and then
we introduce the structure of \package{TorchAudio} and the functionalities available
in Section~\ref{sec:package}.
Finally, we showcase the empirical study we conducted in Section~\ref{sec:empirical}.


\section{Related work}
\label{sec:rel}

Arguably, modern deep learning applications for audio/speech processing
are mainly developed within either the
\package{numpy}~\cite{harris2020array},
\package{TensorFlow}~\cite{tensorflow2015-whitepaper},
or \package{PyTorch}~\cite{NEURIPS2019_9015} ecosystem.
Users tend to commit to one ecosystem when building their
applications to avoid complicated dependencies and conversions between ecosystems that increase 
the cost of maintenance.
There are many excellent open-source toolkits that provide audio/speech
related building blocks and functionalities within each ecosystem.
For example,
\package{librosa}~\cite{mcfee2015librosa} is built for \package{numpy},
\package{DDSP}~\cite{engel2020ddsp} and
\package{TensorFlowASR}\footnote{\url{https://github.com/TensorSpeech/TensorFlowASR}} are created for \package{TensorFlow}, and
\package{TorchAudio} is designed to work with \package{PyTorch}.
\package{TorchAudio} is the go-to toolkit
for basic audio/speech functionalities inside the \package{PyTorch} ecosystem.

\package{TorchAudio} provides important low-level functionalities
like audio input/output, spectrogram computation, and unified 
interface for accessing dataset.
In the \package{PyTorch} ecosystem, there are many useful audio/speech
toolkits available including
\package{Asteroid}~\cite{Pariente2020Asteroid},
\package{ESPnet}~\cite{watanabe2018espnet},
\package{Espresso}~\cite{wang2019espresso},
\package{fairseq}~\cite{ott2019fairseq},
\package{NeMo}~\cite{kuchaiev2019nemo}.
\package{Pytorch-Kaldi}~\cite{pytorch-kaldi}, and
\package{SpeechBrain}~\cite{ravanelli2021speechbrain}.
These toolkits provide ready-to-use models for various applications,
including
speech recognition,
speech enhancement,
speech separation,
speaker recognition,
text-to-speech, etc.
One common thing with these toolkits is that they all have
\package{TorchAudio} as a dependency so that they do not
have to re-implement the basic operations\footnote{We say a package depends on \package{TorchAudio} if they have \code{import torchaudio} in their repository.}.

\section{Design Principles}
\label{sec:design}

\package{TorchAudio} is designed to provide building blocks for the development of
audio applications within the \package{PyTorch} ecosystem.
The functionalities in \package{TorchAudio} are built to be compatible with
\package{PyTorch} core functionalities like neural network containers
and data loading utilities.
This allows users to easily incorporate functionalities in \package{TorchAudio}
into their use cases.
For simplicity and ease of use, \package{TorchAudio} does not depend on any other
\package{Python} packages except \package{PyTorch}.



We ensure that most of our functionalities satisfy three key properties:
(1) GPU compute capability,
(2) automatic differentiability, and
(3) production readiness.
To ensure GPU compute capability, we implement all computational-heavy operations, such as convolution and matrix multiplication, using GPU-compatible logic.
To ensure automatic differentiability, we perform a gradient test on all necessary functionalities.
Lastly, for production readiness, we make sure that most of
our functionalities are compilable into \textit{TorchScript}\footnote{\url{https://pytorch.org/docs/stable/jit.html}}.
TorchScript is an intermediate representation that can be saved from Python code, serialized, and then later loaded by a process where there is no \package{Python} dependency, such as systems where only C++ is
supported\footnote{\url{https://github.com/pytorch/audio/tree/main/examples/libtorchaudio}} and 
mobile platforms including
Android\footnote{\url{https://github.com/pytorch/android-demo-app/tree/master/SpeechRecognition}} and 
iOS\footnote{\url{https://github.com/pytorch/ios-demo-app/tree/master/SpeechRecognition}}.

Since users depend on \package{TorchAudio} for foundational building blocks,
we must maintain its stability and availability to a wide range of platforms, particularly as its user base grows.
We support all platforms that \package{PyTorch} supports, which includes
major platforms (Windows, MacOS, and Linux) with major \package{Python} versions
from $3.6$ to $3.9$.
The features are classified into three release status: \textit{stable},
\textit{beta}, and \textit{prototype}.
Once a feature reaches the stable release status, we aim to maintain backward compatibility whenever it is possible.
Any backward-compatibility-breaking change on stable features can be released only
after two release cycles have elapsed from when it was proposed.
On the other hand, the beta and prototype features can be less stable in terms
of its API, but they allow users to benefit from accessing newly implemented
features earlier.
The difference between beta and prototype is that prototype features are generally
even less stable and will only be accessible from the source code or
nightly build but not from \package{PyPI} or \package{Conda}.
Before the release of a new version,
we use release candidates to collect
user feedback before the official release.

We apply modern software development practices to ensure the quality of the codebase.
For code readability, the code linting complies with PEP-8 recommendations,
with only small modifications.
All developments are conducted openly on GitHub and
all functions are thoroughly documented using 
\package{Sphinx}\footnote{\url{https://www.sphinx-doc.org/en/master/}}.
We also have implemented more than $6000$ test cases using the
\package{pytest}\footnote{\url{https://pytest.org/}} framework and use
CircleCI\footnote{\url{https://circleci.com/}}
for continuous integration testing.

We strive to be selective about the features we implement.
We aim to maintain only the most essential functionalities to keep the design of the package lean.
This way, it is easier to keep each releases stable and the maintenance cost down.
We provide models that researchers would employ as baselines for comparison.
For example, Tacotron2~\cite{shen2018natural}
is usually used as the baseline when developing text-to-speech systems.
Therefore, we included the implementation of Tacotron2.

For sophisticated downstream applications, we provide examples and tutorials\footnote{\url{https://pytorch.org/tutorials/}}
on important downstream applications, therefore,
it is easy for users to adapt \package{TorchAudio} to various applications and use cases.



\section{Package Structure and Functionalities}
\label{sec:package}

Here, we provide an overview of the functionalities and structure of
\package{TorchAudio}.
\package{TorchAudio} supports all four categories of functionalities
mentioned previously, including audio input/output (I/O), audio/speech datasets,
audio/speech operations, and audio/speech models.

\subsection{Audio input/output (I/O)}
Audio I/O is implemented under the \code{backend} submodule to provide
a user-friendly interface for loading audio files into \package{PyTorch}
tensors and saving tensors into audio files.
We ported \texttt{SoX}\footnote{\url{https://sourceforge.net/projects/sox/}} into
\package{TorchAudio} and made it torchscriptable (meaning it is production-ready and
can be used for on-device stream ASR support) for this purpose.
Optionally, \package{TorchAudio} also provides interfaces for different backends such as
\texttt{soundfile} and \texttt{kaldi-io}\footnote{\url{https://github.com/vesis84/kaldi-io-for-python}}.

\subsection{Audio/speech datasets}
The access to audio/speech datasets is implemented under the
\code{dataset} submodule.
The goal of this submodule is to provide a user-friendly interface for accessing commonly used datasets.
Currently, we support $11$ datasets, including
Librispeech~\cite{panayotov2015librispeech},
VCTK~\cite{yamagishi2019cstr},
LJSpeech~\cite{ljspeech17}, etc.
These datasets are designed to greatly simplify data pipelines and are built to be compatible with
the data loading utilities provided in \package{PyTorch},
i.e. \code{torch.utils.data.DataLoader}\footnote{\url{https://pytorch.org/docs/stable/data.html}}.
This allows users to access a wide range of off-the-shelf features
including customizing data loading order, automatic batching,
multi-process data loading, and automatic memory pinning.

\subsection{Audio/speech operations}
We implemented common audio operations under three submodules --
\code{functional},
\code{transform}, and
\code{sox\_effects}.

\paragraph*{\code{functional}.}
In this submodule, we provide $49$ different commonly used operations.
The operations can be further categorized
into four categories including general utility,
complex utility, filtering and feature extraction.
In general utilities, we provide utilities such as
creating of the discrete cosine transformation matrix.
In complex utilities, we provide functions such as
computing the complex norms.
For filtering, we include things like bandpass filters.
Finally, for feature extraction, we have algorithms such
as the computation of spectral centroid.

\paragraph*{\code{transform}.}
There are $26$ different transforms implemented under the
submodule -- \code{torch.nn.Module}.
The objects here are designed to work as neural network building blocks.
They are designed to interface with \package{PyTorch}'s neural network containers
(i.e. \code{torch.nn.Sequential}, \code{torch.nn.ModuleList}, etc.)
As such, they can be seamlessly integrated with all the neural network features in \package{PyTorch}.
The functionalities implemented in this submodule include
Spectrogram/InverseSpectrogram,
mel-frequency cepstrum coefficients (MFCC),
minimum variance distortionless response~(MVDR) beamforming, RNN-Transducer Loss, etc.

\paragraph*{\code{sox\_effects}.}
\package{SoX} is a popular command line program that provides a many
audio processing functionality.
To make these functionalities more accessible, we modified from the source code
of \package{Sox} version 14.4.2 and ported it into \package{TorchAudio}.
This module is fully torchscriptable and supports
$58$ different sound effects, such as resampling, pitch shift, etc.

\subsection{Machine learning models}
We implement various canonical machine learning models
in \code{torchaudio.models} across a wide range of audio-related applications.
For speech recognition, we have 
DeepSpeech~\cite{hannun2014deep}, HuBERT~\cite{hsu2021hubert}, Wav2letter~\cite{collobert2016wav2letter}, and
Wav2Vec 2.0~\cite{baevski2020wav2vec}.
For speech separation, we have
Conv-TasNet~\cite{luo2019conv}.
For text-to-speech and neural vocoder, we have
Tacotron2~\cite{kalchbrenner2018efficient}
paired with
WaveRNN~\cite{kalchbrenner2018efficient}.
These models are selected to support basic downstream tasks.

\begin{figure*}[!ht]
    \centering
    \includegraphics[width=.9\textwidth]{"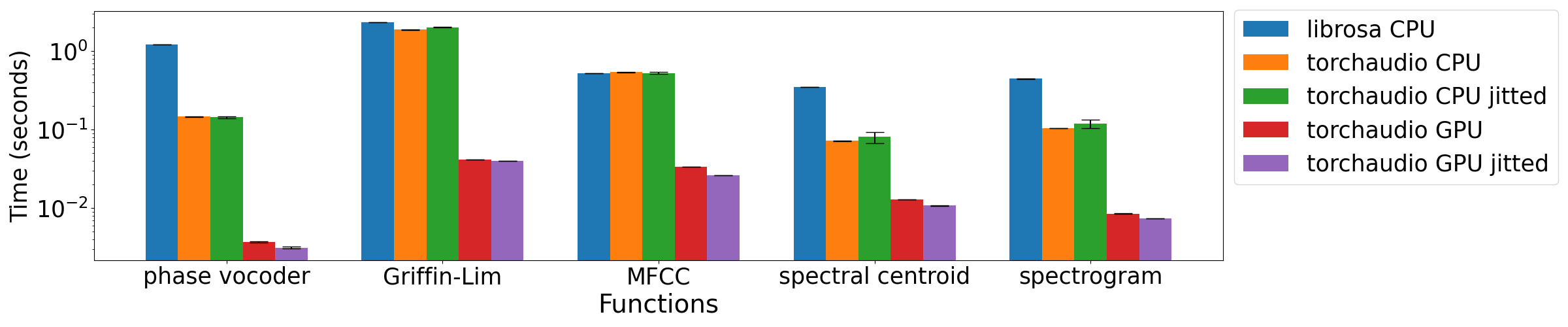"}
    \vspace{-1.5em}
    \caption{The mean and standard error of the running time over five repeated experiments.
      We compare over five different audio processing functions
      using the implementations from \package{TorchAudio} (including  just-in-time compiled (jitted) and GPU accelerated versions) and \package{librosa}.}
    \label{fig:function_bench}
\end{figure*}

\section{Empirical evaluations}
\label{sec:empirical}

This section provides a comparison of our implementations with others.
We present the performance benchmarks for five
implementations of audio/speech operations and three implementations of machine learning models.
We benchmark the audio/speech operations using run time as the main performance metric.
For machine learning models, the performance measures are chosen based on the task
they are performing.
The experiments are conducted on Amazon AWS p4d.24xlarge instance with
NVIDIA A100 GPUs.
The implementation for the empirical evaluation can be found at
\url{https://github.com/yangarbiter/torchaudio-benchmark}.

\subsection{Audio/Speech Operations}

The pioneering \package{librosa}~\cite{mcfee2015librosa} is arguably
one of the most commonly used toolkits in the \package{Python} community.
We select five operations that are implemented in both
\package{TorchAudio} and \package{librosa} and compare their run time.
For \package{librosa}, we use version 0.8.0 installed from \package{PyPI}.
The inputs are all floating point data type.
We consider five different operations including 
phase vocoder, Griffin-Lim algorithm, MFCC, spectral centroid, and spectrogram.
For MFCC, spectral centroid, spectrogram, we measure the run time of $100$ runs,
and for Griffin-Lim algorithm, phase vocoder, we measure the run time of $10$ run.
Operations in \package{TorchAudio}
can be compiled into torchscript during runtime and can be accelerated
with a GPU, thus, we also include the just-in-time compiled (jitted) and GPU 
accelerated versions of the \package{TorchAudio} operations.

The benchmark results are shown in Figure~\ref{fig:function_bench}.
We have two findings.
First, we can see that in terms of CPU operations, \package{TorchAudio} is able to have
even or slightly better speed.
In addition, the support of GPU operation allows \package{TorchAudio} to take advantage of
hardware for acceleration.
Second, we found that for the jitted version,
it runs slightly slower on CPU, but slightly faster on GPU.
However, this difference is marginal.


\subsection{Audio/Speech Applications}

Here, we benchmark the performance of three models, 
including WaveRNN~\cite{kalchbrenner2018efficient},
Tacotron2~\cite{shen2018natural}, and
and Conv-TasNet~\cite{luo2019conv}.
We compare each model independently with a popular
open-source implementation available online.





\paragraph*{WaveRNN.}

\begin{table}[h]
    \centering
    \setlength{\tabcolsep}{4pt}
    \begin{tabular}{l|ccc}
\toprule
                       &   PESQ ($\uparrow$) & STOI ($\uparrow$) & MOS ($\uparrow$) \\
\midrule
fatchord's WaveRNN\tablefootnote{\url{https://github.com/fatchord/WaveRNN}}
                                   & 3.43 & 0.96 & 3.78 $\pm$ 0.08 \\
\package{TorchAudio} WaveRNN       & 3.68 & 0.97 & 3.88 $\pm$ 0.08 \\
\midrule
\package{Nvidia} WaveGlow          & 3.78 & 0.97 & 3.72 $\pm$ 0.05 \\
\package{TorchAudio} WaveRNN       & 3.63 & 0.97 & 3.86 $\pm$ 0.05 \\
\midrule
ground truth                       & - & - & 4.09 $\pm$ 0.05 \\
\bottomrule
\end{tabular}


    \caption{The average PESQ, STOI, and MOS of the vocoders.
    In addition, we also show the standard error of the mean for the MOS metric
    due to larger variance.
    These numbers are higher the better.
    The \package{TorchAudio} WaveRNN on the upper part of the table is trained
    on the dataset split by fatchord's WaveRNN implementation, and
    the \package{TorchAudio} WaveRNN on the lower part of the table is trained
    on the dataset split from Nvidia.
    }
    \label{tab:vocoder_perf}
\end{table}

For WaveRNN, we consider the task to reconstruct a waveform from the corresponding mel-spectrogram.
We measure the reconstruction performance with
the wide band version of the \textit{perceptual evaluation of speech quality~(PESQ)}, and
\textit{short-time objective intelligibility~(STOI)}
on the LJSpeech~\cite{ljspeech17} dataset.
When evaluating with PESQ, we resample the waveform to $16000$ Hz 
and measure the reconstruction performance with a publicly available
implementation\footnote{\url{https://github.com/ludlows/python-pesq}}.
We compare the performance of our model with
the popular WaveRNN implementation from
fatchord\footnote{\url{https://github.com/fatchord/WaveRNN}} as well as
the WaveGlow~\cite{prenger2019waveglow} from
Nvidia\footnote{\url{https://github.com/NVIDIA/DeepLearningExamples}}
The results are shown in Table~\ref{tab:vocoder_perf}.
Note that the pretrained models for fatchord and Nvidia are trained on a different dataset
split, therefore we train and evaluate our WaveRNNs on each of their datasets, respectively.

In addition, we also consider a subjective evaluation metric, 
the mean opinion score (MOS).
For the comparison with fatchord's WaveRNN, we use the full test set, which consists of 
$50$ audio samples; for the comparison with Nvidia's WaveGlow, 
we randomly selected $100$ audio examples from the test set to generate the audio sample.
These generated samples are then sent to Amazon’s Mechanical Turk,
a human rating service, to have human raters score each audio sample.
Each evaluation is conducted independently from each other, so the
outputs of two different models are not directly compared when raters are
assigning a score to them.
Each audio sample is rated by at least $4$ raters on a scale
from $1$ to $5$ with $1$ point increments, where $1$ is lowest
means that the rater perceived the audio as unlike a human speech,
and $5$ means that the rater perceived the audio as similar to a human speech.

The results in Table~\ref{tab:vocoder_perf} show that our implementation is able to achieve similar performance comparing with both fatchord's and Nvidia's models.
This verifies the validity of our implementation.
In addition, our implementation also utilizes the latest 
DistributedDataParallel\footnote{\url{https://pytorch.org/tutorials/intermediate/ddp_tutorial.html}}
for multi-GPU training, which gives additional running time benefit over fatchord's implementation.

\paragraph*{Tacotron2.}
For Tacotron2, we measure the performance with the
\textit{mel cepstral distortion~(MCD)} metric, and we compare our implementation with
Nvidia's implementation\footnote{\url{https://github.com/NVIDIA/DeepLearningExamples}}.
The evaluation is conducted on the LJSpeech dataset.
We train our implemented Tacotron2 using the training testing split from 
Nvidia's repository and compare it with Nvidia's pretrained model.
We train our Tacotron2 similarly to the default parameter provided in 
Nvidia's repository, which has $1500$ epochs, an initial learning rate of $10^{-3}$,
a weight decay of $10^{-6}$, clips the norm of the gradient to $1.0$, and uses
the Adam optimizer.
Table~\ref{tab:tacotron2_perf} shows that
the difference in MCD between 
the two implementations is very small (around $1\%$), which verifies the validity of our implementation.

\begin{table}[h]
    \centering
    \begin{tabular}{l|c}
\toprule
                       &  MCD ($\downarrow$) \\
\midrule
Nvidia\tablefootnote{\url{https://github.com/NVIDIA/DeepLearningExamples/tree/master/PyTorch/SpeechSynthesis/Tacotron2}} &   2.294 \\
\package{TorchAudio}   &   2.305 \\
\bottomrule
\end{tabular}
    \caption{The MCD of different implementation of Tacotron2.
        The numbers are lower the better.}
    \label{tab:tacotron2_perf}
\end{table}

\paragraph*{Conv-TasNet.} We train the Conv-TasNet model on the \textit{sep\_clean} task of Libri2Mix dataset. The sample rate is 8000 Hz. We follow the same model configuration and training strategy in the Asteroid training pipeline: 1) We use the negative value of scale-invariant scale-to-distortion ratio (Si-SDR) as the loss function with a permutation invariant training (PIT) criterion; 2) We clip the gradient with a threshold of 5; 3) We use 1e-3 as the initial learning rate and halve it if the validation loss stops decreasing for 5 epochs. We evaluate the performance by Si-SDR improvement~(Si-SDRi) and signal-to-distortion ratio improvement~(SDRi). Table~\ref{tab:convtasnet_ref} shows that our implementation slightly outperforms that in Asteroid on both Si-SDRi and SDRi metrics.
\begin{table}[h]
    \centering
    \begin{tabular}{l|c|c}
\toprule
                       &  Si-SDRi (dB) ($\uparrow$) & SDRi (dB) ($\uparrow$)\\
\midrule
Asteroid\tablefootnote{\url{https://github.com/asteroid-team/asteroid/tree/master/egs/librimix/ConvTasNet}} &    14.7 & 15.1\\
\package{TorchAudio}   &   15.3 & 15.6 \\
\bottomrule
\end{tabular}
    \caption{The scale-invariant signal-to-distortion ratio~(Si-SDR) and SDR improvements of Conv-TasNet (numbers are higher the better).\vspace{-1em}}
    \label{tab:convtasnet_ref}
\end{table}

\section{Conclusion}
This paper provides a brief summary of the \package{TorchAudio} toolkit.
With increasing number of users, this project is under active development
so that it can serve
our goal to accelerate the development and deployment of audio-related 
machine learning applications.
The roadmap for future work can be found on
our Github page\footnote{\url{https://github.com/pytorch/audio}}.

\section{Acknowledgement}

We thank the release engineers, Nikita Shulga
and Eli Uriegas, for helping with creating releases.
We thank all contributors on Github, including but not limited to 
Bhargav Kathivarapu,
Chin-Yun Yu,
Emmanouil Theofanis Chourdakis,
Kiran Sanjeevan, and
Tomás Osório\footnote{Due to page limit, we only include the name of people who contributed $1000+$
lines as of September 2021.}.
We thank
Abdelrahman Mohamed,
Alban Desmaison,
Andrew Gibiansky,
Brian Vaughan,
Christoph Boeddeker,
Fabian-Robert Stöter,
Jan Schlüter,
Keunwoo Choi,
Manuel Pariente,
Mirco Ravanelli,
Piotr Żelasko,
Samuele Cornell,
Yossi Adi, and
Zhi-Zheng Wu for the meaningful discussions about
the design of \package{TorchAudio}.
We also thank the Meta marketing team that helped with promoting this work.

{
\footnotesize
\bibliographystyle{IEEEbib}
\bibliography{main}
}

\end{document}